\documentstyle[preprint,revtex]{aps}
\begin{document}
%
                \hbox{December 1993}}$ }

\begin{title}
\begin{center}
On the Precision of the Computation of the QCD Corrections to Electroweak
Vacuum Polarizations
\end{center}
\end{title}
\author{M.C. Gonzalez-Garcia, F. Halzen, and R.A. V\'azquez}
\begin{instit}
 Physics Department, University of Wisconsin, Madison, WI 53706
\end{instit}
\thispagestyle{empty}
\begin{abstract}
We demonstrate that the dispersive computation of the threshold enhancements
to heavy quark vacuum polarizations is unstable. Because of the slow
convergence of the dispersion relations the result critically depends on the
intermediate energy region where the non-relativistic approximation, intrinsic
to threshold calculations, is invalid. We discuss other ambiguities precluding
a reliable calculation of the threshold contribution to the vacuum
polarizations. In the absence of a solution prudence should force one to
assign an error to the radiative corrections not far below the level of the
pertubative O($\alpha \alpha_s$) contributions. This may preclude the
extraction of the Higgs mass from precision measurements.
\end{abstract}

\newpage
Measurements of the electroweak parameters are sufficiently precise to require
a quantitative evaluation of the QCD corrections \cite{fanchi}. Diagrams where
a gluon is exchanged between the b,t-quarks inside vacuum polarization loops
cannot be neglected. Whereas the perturbative part of this ${\cal O}(\alpha
\alpha_s)$ gluon correction to the quark loops can be reliably calculated using
either dispersive or straightforward diagrammatic techniques
\cite{gaemers,pert,stuart}, evaluation of the threshold contribution of heavy
quark bound states seems less straightforward. We demonstrate in this paper
that the use of dispersion relations to account for threshold effects is, in
this case, unfortunately intrinsically unstable. The calculational results
critically depend on the behavior of the vacuum polarizations functions at an
intermediate energy where the asymptotic (perturbative) limit is not valid, but
which is sufficiently high to invalidate the threshold resummation.
Alternatively, the result depends critically on the computation of the
threshold effects in an energy region where the non-relativistic approximation,
central to all estimates, is no longer valid.

Although we offer no solution, we find the problem sufficiently important to
expose the unstable nature of the computation in this paper. Conflicting
results in the literature can be readily understood. More importantly, the
ambiguity associated with the computation of the QCD corrections to the vacuum
polarizations is large enough to make any determination of the Higgs mass via
high precision measurements of electroweak parameters questionable in the
absence of a resolution of the problem described in what follows.

Computation of QCD corrections to the vacuum polarization can be performed via
once-subtracted dispersion relations. As usual, the subtraction is not unique.
Two subtraction prescriptions appear in the literature by Sirlin and Kniehl(SK)
\cite{sirlin_1},
\begin{equation}
\Pi^{V,A}(s,m_1,m_2)=\frac{1}{\pi}\left[
\int_{(m_1+m_2)^2}^{\Lambda^2}ds'\frac{Im \Pi^{V,A}(s',m_1,m_2)}
{s'-s-i\epsilon} +
\int_{(m_1+m_2)^2}^{\Lambda^2} ds' Im \lambda^{V,A}(s',m_1,m_2) \right]
\label{bernd}
\end{equation}
and Gaemers et al.(G) \cite{gaemers}
\begin{equation}
\Pi^{V,A}(s,m_1,m_2)=\frac{1}{\pi}\left[
\int_{(m_1+m_2)^2}^{\Lambda^2}ds'\frac{Im \Pi^{V,A}(s',m_1,m_2)}
{s'-s-i\epsilon} -\frac{1}{2} \sum_{i=1}^{2}
\int_{4m_i^2}^{\Lambda^2} ds'\frac{Im \Pi^{V}(s',m_i,m_i)}
{s'-s-i\epsilon}\right],
\end{equation}
where
$\Pi^{V,A}$ and $\lambda^{V,A}$ are the $g_{\mu\nu}$ and $q_\mu q_\nu$
coefficients of the vacuum polarization tensors for the vector and axial
currents, {\it i.e.}
\begin{equation}
\Pi_{\mu\nu}^{V,A}(q,m_1,m_2)=\Pi^{V,A}(s,m_1,m_2)g_\mu\nu
+\lambda^{V,A}(s,m_1,m_2) q_\mu q_\nu
\end{equation}
Perturbatively, the functions $\Pi$ and $\lambda$ can be expanded in the
coupling constant:
\begin{equation}
\Pi(s)=\Pi^0(s)+\frac{\alpha_s}{\pi} \Pi^1(s) + ...
\label{def}
\end{equation}

Both subtractions yield identical results for the perturbative gluon correction
to quark loops in the gauge bosons self energies. This result also agrees with
direct computation of the 2-loop diagrams for a constant value of $\alpha_s$.
This was first demonstrated for the case of a massless bottom quark. Recently
the equivalence between the SK dispersion and the dimensional regularization
evaluation of $\Delta\rho$ was established to O($\alpha\alpha_s$)
\cite{djouadi} for a finite $m_b$. We explicitly demonstrated that both
subtraction prescriptions yield the same result for a finite $m_b$ for $\Delta
r$ and $\Delta\rho$.

Although the choice of subtraction is ambiguous, arguments can be advanced
supporting the scheme proposed by Kniehl and Sirlin \cite{sirlin_2}. They
demonstrated that the gluon corrections to the process $H^0 \rightarrow l^+
l^-$ are different for the two dispersion relations at the $O( \alpha
\alpha_s)$ level. The result of a direct computation of the diagrams using
dimensional regularization agrees with the first approach, Eq. \ref{bernd}. In
the following we will therefore use the SK subtraction.

The dispersive approach is, in principle, most suitable to incorporate the
$t\bar t$ threshold enhancement into the vacuum polarization functions since
such effects are usually expressed in terms of contributions to the absorptive
parts. They are calculated below threshold by summing over the $t \bar t$
resonances,
\begin{equation}
\begin{array}{l}
\frac{1}{s}Im \Pi^{V}(s,m_t,m_t)=
N_c (1-4 C_F \frac{\alpha_s}{\pi}){\displaystyle \sum_n} \frac{|R_n(0)|^2}
{ M_n} \delta(s-M_n^2) \\-Im \lambda^{A}(s,m_t,m_t)=
N_c (1-3 C_F \frac{\alpha_s}{\pi}) {\displaystyle \sum_n} \frac{|R_n(0)|^2}
{ M_n} \delta(s-M_n^2)
\end{array}
\end{equation}
where $R_n$ is the wave function of the S-wave nth resonance $t\bar t$ state
with mass $M_n$.

Above threshold there are a variety of equivalent ways to introduce the
threshold effects. They may be estimated using the Green's function by solving
the Schr\"odinger equation \cite{green}. The Green function is readily related
to the vacuum polarization functions:
\begin{equation}
\begin{array}{l}
\frac{1}{s}Im \Pi^{V}(s,m_t,m_t)=
N_c (1-4 C_F \frac{\alpha_s}{\pi})
\frac{1}{2 m_t^2} Im G(0,0;\sqrt{s}-2m_t+i\Gamma_t)\\
-Im \lambda^{A}(s,m_t,m_t)=
N_c (1-3 C_F \frac{\alpha_s}{\pi})
\frac{1}{2 m_t^2} Im G(0,0;\sqrt{s}-2m_t+i\Gamma_t).\\
\end{array}
\end{equation}
$G(0,0;\sqrt{s}-2m_t+i\Gamma_t)$ is the Green's function evaluated at the
origin and $\Gamma_t$ is the top width. Although the Green's function must, in
general, be evaluated numerically, for a Coulomb potential the Schr\"odinger
equation is solvable and we obtain the well-known resummation factor
\begin{equation}
\begin{array}{l}
\frac{1}{s}Im \Pi^{V}(s,m_t,m_t)= N_c (1-4 C_F \frac{\alpha_s}{\pi})
\frac{\displaystyle \frac{4}{3}\frac{\pi\alpha_s}{\beta}}
{\displaystyle 1-\exp(-\frac{4}{3}\frac{\pi\alpha_s}{\beta})}
\frac{1}{s}Im \Pi_0^{V}(s,m_t,m_t) \\
-Im \lambda^{A}(s,m_t,m_t)= N_c (1-3 C_F \frac{\alpha_s}{\pi})
\frac{\displaystyle \frac{4}{3}\frac{\pi\alpha_s}{\beta}}
{\displaystyle 1-\exp(-\frac{4}{3}\frac{\pi\alpha_s}{\beta})}
\frac{1}{s}Im \Pi_0^{V}(s,m_t,m_t)   \\
\label{coulombic}
\end{array}
\end{equation}
These equations give the correct behavior of the imaginary parts of the vacuum
polarization functions close to threshold. The factor $(1-  C_F \alpha_s/\pi)$
incorporates the next term in the threshold expansion $ (\alpha_s/\beta)^n$,
where $\beta$ is the relative velocity between the $t \bar t$ pair. We show in
figure \ref{potential} the solutions of the Schr\"odinger equation for the
potential J of Ref. \cite{igi_ono} for three different values of
$\Lambda^4_{QCD}$. Also shown are the perturbative results for three values of
the argument of $\alpha_s$ ($\alpha_s(m_{t}/4),\alpha_s(m_{t}), \alpha_s(4
m_{t})$). We here should draw attention to the fact that the threshold
resummation is valid only near threshold, {\it i.e.}
\begin{equation}
\beta = \left( 1 - {4m_t^2\over s}\right)^{1/2} \ll 1 \,.
\end{equation}

At this point the calculation is, in principle, straightforward. One calculates
the threshold behavior of the quarks in the loop at low $\beta$ and
subsequently ``forces'' the result to smoothly join the perturbative answer
\cite{stuart}. The latter indeed represents the correct answer far above
threshold. The value where this junction occurs we denote by $\beta_{\rm cut}$.
The contribution above threshold to the real part of the polarization functions
is then calculated by
\begin{equation}
\Pi(s')=\frac{1}{\pi} \int^{s_{cut}}_{4 m_{t}^2} ds \;
\frac{Im \Pi_{\rm thres}(s)-Im \Pi_{\rm pert}(s)}{(s-s')}
\label{thres}
\end{equation}
where $s_{\rm cut}=s(\beta_{\rm cut})$. One must realize that, no matter what
approach one follows, an equation like (\ref{thres}) holds. E.g., Yndur\'ain
introduces it explicitly in eqn. (11) of Ref. \cite{yndurain}. In the paper by
Sirlin and Kniehl \cite{sirlin_2}, they simulate the threshold behavior by
introducing a running $\alpha_s$ with an argument which is singular at
threshold
\begin{equation}
Im \Pi_{\rm thres}(s)= Im \Pi^0(s) +\frac{\alpha_s(x p^2)}{\pi} Im \Pi^1(s)
\label{thres_bernd}
\end{equation}
where $p=\sqrt{\frac{s}{4}-m_t^2}$ is the three momentum of the $t \bar t$ pair
and $x$ is a number of order 1. $Im \Pi^{0,1}$ are defined by eq. (\ref{def}).
The threshold contribution is then obtained by integrating the function
(\ref{thres_bernd}) up to the matching point $x p^2= m_{t}^2$. In this way they
fit the threshold function by the running of $\alpha_s$ at low $\beta$ and, at
high energy, they match the perturbative expression.

It is clear that these procedures are far from unique. We will see that the
results, unfortunately depend critically on these arbitrary procedures. We
illustrate this further by using the elegant technique proposed by Voloshin
\cite{voloshin}. Voloshin has obtained an interpolating formula between the
threshold behavior and the asymptotic perturbative result for the production of
$\tau^+ \tau^-$ from $e^+ e^-$ \cite{voloshin}. This formula correctly accounts
for the limiting behavior at both ends ($\beta \rightarrow 0,1$) and, smoothly
interpolates between them. It is straightforward to rewrite it for a heavy
quark threshold. Using the Coulombic resummation (\ref{coulombic}) his result
can be written as
\begin{equation}
Im \Pi(s)= F_t(s) \; Im \Pi^0(s) \; (1+X(s))
\label{thres_volo}
\end{equation}
where $F_t=\frac{\displaystyle \frac{4}{3}\frac{\pi\alpha_s}{\beta}}
{\displaystyle 1-\exp(-\frac{4}{3}\frac{\pi\alpha_s}{\beta})}$ is the Coulomb
threshold factor, $Im \Pi^0$ is the zero order, one loop, vacuum polarization
function and $X$ can be written as
\begin{equation}
\begin{array}{ll}
X(\beta)=& \frac{\displaystyle \alpha_s}{\displaystyle \pi \beta}
\left\{
 (1+\beta^2) \left[ \ln \frac{1+\beta}{2}
\ln \frac{1+\beta}{1-\beta} +2 Li_2(\frac{1-\beta}{1+\beta}) -\frac{\pi^2}{3}
 +2 Li_2(\frac{1+\beta}{2})-2 Li_2(\frac{1-\beta}{2})\right.\right. \\
& \left. -4 Li_2(\beta)+Li_2(\beta^2) \right]
+\left[ \frac{11}{8} (1+ \beta^2) -3 \beta +
\frac{1}{2} \frac{\beta^4} {3-\beta^2} \right] \ln \frac{1+\beta}{1-\beta}
+ 6 \beta \ln \frac{1+\beta}{2}\\
&\left. - 4 \beta \ln \beta  +\frac{3}{4} \beta \frac{5-3\beta^2}{3-\beta^2}
\right\}.
\end{array}
\end{equation}
Eq. (\ref{thres_volo}) incorporates the radiative corrections with correct
$\beta \rightarrow 0,1$ limits. In the above equation, we will parametrize the
scale of $\alpha_s$ to be
\begin{equation}
\mu^2=x p^2+  \beta \; \; (\mu^2_{\rm pert} - x p^2)
\end{equation}
where $\mu_{\rm pert}$ is the perturbative scale, and $x$ is a number of order
1. This way we reproduce indeed the threshold correction at low $\beta$ and the
perturbative result at high $\beta$. In figure \ref{imaginarios} we plot the
perturbative imaginary vacuum polarization function for $\mu^2_{\rm pert}=
m_t^2$, the threshold function as given by eq. (\ref{thres_bernd}) and the
Voloshin result of eq. (\ref{thres_volo}) for $x=1/4,1,4$. As expected, the
matching point is a function of $x$, a parameter undetermined to this order.
The crucial point is that the dependence of the result on the value of
$\beta_{\rm cut}$ is highly unstable. Numerically $\beta_{\rm cut}$ is not
small. In the paper of Yndur\'ain, he writes $\beta_{\rm cut}=\pi /\sqrt 2 C_F
\alpha_s$ which is of the order of 0.3. In the work by Sirlin and Kniehl, they
use $x=4$ which gives $\beta_{\rm cut} = \sqrt(1- 4/5) = 0.45 $. As will be
illustrated next, the threshold contribution to the electroweak parameters
obtained using eq. (\ref{thres}) not only depend strongly on $\beta_{\rm cut}$,
one might question the use of the non-relativistic approximation at such large
values of the velocity. We have also investigated relativistic solutions for
the Coulomb potential \cite{durand}. Again, no stable result was found.

We next calculate the threshold contribution of the $t \bar t$ pair to the
electroweak parameter $\Delta \rho$ and $\Delta r$. In figures
\ref{deltar}-\ref{deltarho} we plot the contribution to these parameters as a
function of the ``matching point'' for the approach of Sirlin and Kniehl.
We repeat the calculation using the Voloshin interpolation formula. We present
the results for three values of $x$, $x=1/4,1,4$. In tables \ref{tabla_r} and
\ref{tabla_rho} we show  the order zero and first order contributions to these
parameters as well as the range of threshold contributions obtained using the
two methods. As can be seen from the figures the actual value of the
contribution depends strongly on the value of $\beta_{\rm cut}$ or,
alternatively on the value of $x$.

We conclude that
\begin{itemize}
\item  there is a wide range of results which depend on an arbitrary parameter
which is not controlled to this precision of perturbation theory,
\item the threshold enhancement is potentially large. As the calculation turns
out to be unstable it is impossible to make any arguments in favor of small or
large results. The uncalculable threshold contribution is possibly as large as
the total Higgs contribution to $\Delta r$, which varies from $-2.78\times
10^{-3}$ to $1.1\times10^{-2}$  depending on the value of the Higgs mass
(0.1-1~TeV),
\item large values of the threshold enhancement do correspond to large values
of $\beta_{\rm cut}$ and one must conclude that the result, in such a case, is
suspect. The essential problem is however that the dispersion relations
converge slowly and the results therefore critically depend on the threshold
enhancement in a region where the low $\beta$ approximation is no longer valid.
The values for $\beta_{\rm cut}$ are high for all examples.
\end{itemize}
The discussion has so far ignored other challenges in performing a reliable
calculation
\begin{itemize}
\item threshold corrections have only been calculated in the abelian limit,
\item the calculation is unreliable far below threshold,
\item the subtraction is not unique. The two subtraction schemes discussed here
give different results which are roughly opposite in sign. The origin of the
problem is the large mass splitting between top and bottom. We have been
unsuccessful in tracing the discrepancy between both dispersion relations.
\end{itemize}

Obviously the problem discussed here is urgent. In the absence of a solution
prudence will force one to essentially assign errors to the vacuum
polarizations not far below the level of the leading QCD correction. This
precludes extracting any information on the Higgs mass and degrades
significantly the indirect information on the top mass, especially in the case
of a heavy top.

\acknowledgments
We gratefully acknowledge discussions with B.A. Kniehl, M.L. Stong and  M.B.
Voloshin. We would like to thank M.J. Strassler for providing us  with the
numerical solutions for the Schr\"odinger equation. This work is supported in
part by the U.S.~Department of Energy under contract No.~DE-AC02-76ER00881, and
in part by the University of Wisconsin Research Committee with funds granted by
the Wisconsin Alumni Research Foundation.

\begin{table}
\caption{Perturbative and threshold contributions to $\Delta r$. The O($\alpha
\alpha_s$) was computed for $\Lambda^4_{QCD}=400$ MeV and the scales
$\mu^2_{\rm pert}=m_t^2/4,m^2_t $ (central value) $,4m^2_t$. The  contributions
below threshold are given for the J potential for $\Lambda^4_{QCD}=100-500$
MeV. The maximum contribution above threshold is quoted for $x=4$ in
Eq.(\ref{thres_bernd}).}
\label{tabla_r}
\begin{displaymath}
\begin{array}{||c|c|c|c|c||}
\hline
\hline
m_{t} & \mbox{O($\alpha$)} & \mbox{O($\alpha \alpha_s$)}\times 10^3 &
\mbox{below threshold}\times 10^3 & \mbox{above threshold}\times 10^3 \\
\hline
120 & -0.045 & 2.10\begin{array}{l}+0.24\\[-0.4cm]-0.18\end{array}
&0.15 - 0.34 & 0 -0.94  \\
160 & -0.059 & 3.04\begin{array}{l}+0.33\\[-0.4cm]-0.25\end{array}
& 0.28-0.58 & 0 - 1.5\\
200 & -0.076 & 4.32\begin{array}{l}+0.46\\[-0.4cm]-0.34\end{array}
&0.38-0.12 &  0- 2.1 \\
\hline
\hline
\end{array}
\end{displaymath}
\end{table}

\begin{table}
\caption{Same as previous table for $\Delta \rho$.}
\label{tabla_rho}
\begin{displaymath}
\begin{array}{||c|c|c|c|c||}
\hline
\hline
m_{t} & \mbox{O($\alpha$)} & \mbox{O($\alpha \alpha_s$)}\times 10^4 &
\mbox{below threshold} \times 10^4& \mbox{above threshold}\times 10^4 \\
\hline
120 & 0.0043 & -4.56\begin{array}{l}+0.39\\[-0.4cm]-0.52\end{array}
&-(0.43- 0.96) & -(0-2.8) \\
160 & 0.0077 & -7.96\begin{array}{l}+0.82\\[-0.4cm]-0.66\end{array}
&-(0.81- 1.7) & -(0-4.0) \\
200 & 0.012 & -11.8\begin{array}{l}+0.93\\[-0.4cm]-1.2\end{array}
&-(1.1 - 2.4) & -(0-6.3) \\
\hline
\hline
\end{array}
\end{displaymath}
\end{table}

\figure{ Contributions to the imaginary part of the vacuum polarization
function from O($\alpha$)+O($\alpha\alpha_s$) radiative corrections (dotted
lines) and from the threshold enhancement given by the solution of the
Schr\"odinger  equation using the Igi-Ono potentials (full lines). Dotted lines
are  the perturbative functions for $\mu^2_{\rm pert}=m_t^2/4,m^2_t,4m^2_t$
(from to bottom) and  $\Lambda^4_{QCD pert}=400$ MeV. Continues lines are the
solutions of the Igi-Ono potentials for $\Lambda^4_{QCD}=100,300,500$ MeV (from
bottom to top).
\label{potential}}
\figure{Imaginary part of the vacuum polarization function for $m_t=160$ GeV.
Full line represents the perturbative result for  $\mu^2_{\rm pert}=m_t^2$ and
$\Lambda^4_{QCD pert}=400$ MeV.  Dotted lines are the interpolating formula by
Voloshin with $x=1/4,1,4$  (top to bottom). Dashed lines are the perturbative
result with running $\alpha_s$ as used by Sirlin and Kniehl with the same
values of $x$.
\label{imaginarios}}
\figure{Threshold contribution to $\Delta\rho$ for $m_t=160$ GeV as a function
of $\beta_{\rm cut}$ for two approaches discussed in the text. Dashed lines
correspond to the prediction of Sirlin and Kniehl with  $x=1/4,1,4$. Dotted
lines are the prediction using the Voloshin interpolating formula with the same
values of $x$.
\label{deltar}}
\figure{Same as in figure 3 for $\Delta r$.
\label{deltarho}}

\end{document}